\documentclass[11pt]{article}
\usepackage{amsmath}
\usepackage{amssymb}
\usepackage{amstext}
\usepackage{amsthm}
\usepackage{url}

\usepackage{multirow}
\usepackage{graphicx}

\numberwithin{table}{section}

\numberwithin{equation}{section}

\numberwithin{figure}{section}

\newcommand{\DEF}[1]{\stackrel{\sf def}{#1}}
\newcommand{\FRACTIONTHREE}[3]{#1\{{#2}#3\}}
\newcommand{\FRACTION}[1]{\FRACTIONTHREE{\left}{#1}{\right}}

\def\PARTITION{{\sf partition}}
\def\COMPUTE{{\sf compute}}

\def\PartitionInputFormat{{\sf PartitionInputFormat}}
\def\Mapper{{\sf Mapper}}

\def\SingletonInputFormat{{\sf SingletonInputFormat}}
\def\PartitionMapper{{\sf PartitionMapper}}
\def\Indexer{{\sf Indexer}}
\def\Reducer{{\sf Reducer}}

\long\def\symbolfootnote[#1]#2{\begingroup%
\def\thefootnote{\fnsymbol{footnote}}\footnote[#1]{#2}\endgroup} 

\newcommand{\IGNORE}[1]{}

\begin{document}
\title{
The Two Quadrillionth Bit of $\pi$ is 0!
\\
Distributed Computation of $\pi$ with Apache Hadoop
}
\author{
{\Large Tsz-Wo Sze} \\
\\
Yahoo!\ Cloud Computing \\
{\tt tsz@yahoo-inc.com}}
\date{October 6, 2010}
\maketitle

\begin{abstract}
We present a new record on computing specific bits of $\pi$,
the mathematical constant,
and discuss performing such computations on Apache Hadoop clusters.
The specific bits represented in hexadecimal are
\[
\text{\tt
0E6C1294 AED40403 F56D2D76 4026265B CA98511D 0FCFFAA1 0F4D28B1 BB5392B8}. 
\]
These $256$ bits end at the $2,000,000,000,000,252^{\text{nd}}$
bit position\symbolfootnote[2]{
When $\pi$ is represented in binary,
we have $\pi=11.0010\;0100\;\underline{0011\;1111}\ldots$
Bit position is counted starting after the radix point.
For example,
the eight bits starting at the ninth bit position are $0011\;1111$ in binary
or, equivalently, {\tt 3F} in hexadecimal.
},
which doubles the position and quadruples the precision
of the previous known record~\cite{Percival2000}.
The position of the first bit is $1,999,999,999,999,997$
and the value of the two quadrillionth bit is 0.

The computation is carried out by a MapReduce program called {\it DistBbp}.
To effectively utilize available cluster resources
without monopolizing the whole cluster,
we develop an elastic computation framework
that automatically schedules computation slices,
each a DistBbp job,
as either map-side or reduce-side computation
based on changing cluster load condition.
We have calculated $\pi$ at varying bit positions and precisions,
and one of the largest computations took 23 days of wall clock time
and 503 years of CPU time on a 1000-node cluster.
\end{abstract}
\section{Introduction}
The computation of the mathematical constant $\pi$
has drawn a great attention from mathematicians and computer scientists
over the centuries
\cite{Berggren2004, Knuth1997}.
The computation of $\pi$ also serves as
a vehicle for testing and benchmarking computer systems.
There are two types of challenges,
\renewcommand{\labelenumi}{(\roman{enumi})}
\begin{enumerate}
\item computing the first $n$ digits of $\pi$, and
\item computing only the $n^{\text{th}}$ bit of $\pi$.
\end{enumerate}
\renewcommand{\labelenumi}{\arabic{enumi}.}
%
%
%
%
In this paper,
we discuss our experience on computing the $n^{\text{th}}$ bit of $\pi$
with Apache Hadoop (\url{http://hadoop.apache.org}), 
an open-source distributed computing software.
To the best of our knowledge,
the result obtained by us,
the Yahoo!\ Cloud Computing Team,
is a new world record as this paper being written.

In 1996,
Bailey, Borwein and Plouffe discovered a new formula
(equation (\ref{eqn-bbp})) to compute $\pi$,
which is now called the BBP formula \cite{Bailey1997},
\begin{equation}\label{eqn-bbp}
\pi=\sum_{k=0}^\infty \frac{1}{2^{4k}}
\left( \frac{4}{8k+1}
       - \frac{2}{8k+4}
       - \frac{1}{8k+5}
       - \frac{1}{8k+6}
\right).
\end{equation}
The remarkable discovery leads to the first digit-extraction algorithm
for $\pi$ in base $2$.
In other words,
it allows computing the $n^{\text{th}}$ bit of $\pi$
without computing the earlier bits.
Soon after,
Bellard has discovered a faster BBP-type formula \cite{Bellard1997},
\begin{equation}\label{eqn-bellard}
\pi=\sum_{k=0}^\infty \frac{(-1)^k}{2^{10k}}
\left( \frac{2^2}{10k+1}
       - \frac{1}{10k+3}
       - \frac{2^{-4}}{10k+5}
       - \frac{2^{-4}}{10k+7}
       + \frac{2^{-6}}{10k+9}
       - \frac{2^{-1}}{4k+1}
       - \frac{2^{-6}}{4k+3}
\right).
\end{equation}
He computed 152 bits of $\pi$
ending at the $1,000,000,000,151^{\text{st}}$ bit position in 1997
\cite{Bellard1997b}.
The computation took 12 days with more than 20 workstations
and 180 days of CPU time.
In 1998,
Percival started
a distributed computing project called PiHex
to calculate the five trillionth bit,
the forty trillionth bit 
and the quadrillionth bit of $\pi$ \cite{Percival2000}.
The best result obtained was
64 bits of $\pi$ ending at the $1,000,000,000,000,060^{\text{th}}$ position
in 2000.
The entire calculation took two years
and required 250 CPU years,
using idle time slices of 1734 machines in 56 countries.
The ``average'' computer participating was a 450 MHz Pentium II.
For a survey on $\pi$ computations,
see \cite{Borwein2010}.

The remainder of the paper is organized as follows.
The results are presented in next section.
We discuss the BBP digit-extraction algorithm and our implementation
in Section \ref{sect-bbp algorithm}
and Section \ref{sect-implementation},
respectively.
\section{Results}
We have developed a program called {\it DistBbp},
which uses equation (\ref{eqn-bellard})
to compute the $n^{\text{th}}$ bit of $\pi$ with arbitrary precision arithmetic.
DistBbp employs the MapReduce programming model \cite{Dean2004}
and runs on Hadoop clusters.
It has been used to compute 256 bits of $\pi$
around the two quadrillionth bit position
as shown in Table \ref{tab-2e15 pi bits}.
This is a new record,
which doubles the position and quadruples the precision
of the previous record obtained by PiHex.

%
%
%
%
\begin{table}[h]
\begin{center}
\begin{tabular}{|c|l|}
\hline
\multirow{2}{*}{Bit Position $n$}
& Bits of $\pi$ Starting at The $n^{\text{th}}$ Bit Position \\
& (in Hexadecimal) \\
\hline
\hline
1,999,999,999,999,997 & 
  {\tt 0E6C1294 AED40403 F56D2D76 4026265B CA98511D} \\ 
& {\tt 0FCFFAA1 0F4D28B1 BB5392B8} \\ 
& (256 bits) \\ 
\hline
\end{tabular}
\caption{The 256 bits of $\pi$ starting 
at the $(2\cdot10^{15}-3)^{\text{th}}$ bit position
and ending at the $(2\cdot10^{15}+252)^{\text{nd}}$ bit position.}
\label{tab-2e15 pi bits}
\end{center}
\end{table}
We have also computed the first one billion bits
and the bits at positions $n=10^m+1$ for $m\leq 15$.
Table \ref{tab-pi bits} below shows the results for $13\leq m\leq 15$.
The results for $n<10^{13}$ are omitted
since the corresponding bit values are well-known.
It appears that the results for $13\leq m\leq 14$ are new,
although their computation requirements are not as heavy as the one for $m=15$.
The result for $m=15$ is similar to the one obtained by PiHex
except that the starting positions are slightly different
and our result has a longer bit sequence.
These computations were executed on the idle slices
of the Hadoop clusters in Yahoo!.
The cluster sizes range from 1000 to 4000 machines.
Each machine has two quad-core CPUs
with clock speed ranging from 1.8 GHz to 2.5 GHz.
We have run at least two computations at different bit positions,
usually $n$ and $n-4$,
for each row in Tables \ref{tab-2e15 pi bits} and \ref{tab-pi bits}.
Only the bit values covered by two computations are considered as valid results
in order to detect machine errors and transmission errors.
Table \ref{tab-time} below shows the running time information
for some computations.

\begin{table}[h]
\begin{center}
\begin{tabular}{|c|r|l|}
\hline
\multirow{2}{*}{$m$}
& \multirow{2}{*}{Bit Position $n=10^m+1$}
& Bits of $\pi$ Starting at The $n^{\text{th}}$ Bit Position \\
& & (in Hexadecimal) \\
\hline
\hline
13 & 10,000,000,000,001 & 
    {\tt 896DC3D3 6A09E2E9 29CA6F91 66FBA8DC F000C4A6} \\ 
& & {\tt 4C78723F 814F2EB4 6D417E5A 4337FB1C C2EB474F} \\ 
& & {\tt 74CCD953 94FB7045 3F7B48AE E758BDD2 DD7B1371} \\ 
& & {\tt 0CDB80EF 72B70912 E20281FC 76FD0A10 CDE2ADD8} \\ 
& & {\tt BD5163E1 FC582BFE FB4D8F9A F4A771E8 BA9F0B58} \\ 
& & {\tt C0334D55 297ADDEB 1DACB0EF B572D927 DBDDB68D} \\ 
& & {\tt 858929EA D8} \\ 
& & (1000 bits) \\ 
\hline
14 & 100,000,000,000,001 & 
    {\tt C216EC69 7A098CC4 B9AF60D0 5AE28EA9 36873682} \\ 
& & {\tt D062B83B 52C5C205 CDA35F4D BCD0E9C3 785CBFA7} \\ 
& & {\tt E62401AB B69AF82C CE885230 03D4FC01 7C620B11} \\ 
& & {\tt A94B99F7 4DDE5102 A5142280 46B0055A 636715D3} \\ 
& & {\tt 75CB8BAC 2003BA93 27B731EA 40341861 27419284} \\ 
& & {\tt E3FFE685 480637BF 5C5BAE91 3AFB7EA7 45B4C955} \\ 
& & {\tt 8E2EB177} \\ 
& & (992 bits) \\ 
\hline
15 & 1,000,000,000,000,001 & 
    {\tt 6216B069 CB6C1D36 117099E4 3646A6D4 48D887CC} \\ 
& & {\tt D95CC79A C84E60D2 3} \\ 
& & (228 bits) \\ 
\hline
\end{tabular}
\caption{The bits of $\pi$ starting at the $(10^m+1)^{\text{st}}$ positions
for $13\leq m\leq 15$.}
\label{tab-pi bits}
\end{center}
\end{table}

\begin{table}[h]
\begin{minipage}{\linewidth}
\begin{center}
\begin{tabular}{|r|c|c|c|c|}
\hline
\multirow{2}{*}{Starting Bit Position}
& \multirow{2}{*}{Precision (bits)}
& \multirow{2}{*}{Time Used\footnote{
Note that Time Used is not equivalent to ``cluster time''
since there were other jobs running on the cluster.
}}
& \multirow{2}{*}{CPU\footnote{
Note that the CPUs in a cluster may be slightly different.
}
Time}
& \multirow{2}{*}{Date Completed} \\
& & & & \\
\hline
\hline
          1 & 800,001,000 & 10 days & 19 years & June 23, 2010 \\
\hline
800,000,001 & 200,001,000 &  3 days & 8 years & June 22, 2010 \\
\hline
\hline
          999,999,997 & 1024 & 102 seconds & 51 minutes & June 10, 2010 \\
\hline
        1,000,000,001 & 1024 &  96 seconds & 54 minutes & June 11, 2010 \\
\hline
        9,999,999,997 & 1024 &   2 minutes & 21 hours & June 8, 2010 \\
\hline
       10,000,000,001 & 1024 &   4 minutes & 21 hours & June 8, 2010 \\
\hline
       99,999,999,997 & 1024 &  10 minutes &  12 days & June 6, 2010 \\
\hline
      100,000,000,001 & 1024 &   9 minutes &  11 days & June 6, 2010 \\
\hline
      999,999,999,997 & 1024 & 105 minutes & 121 days & June 7, 2010 \\
\hline
    1,000,000,000,001 & 1024 &  98 minutes & 121 days & June 7, 2010 \\
\hline
    9,999,999,999,997 & 1024 & 10 hours & 4 years & June 2, 2010 \\
\hline
   10,000,000,000,001 & 1024 &  8 hours & 4 years & June 1, 2010 \\
\hline
   99,999,999,999,997 & 1024 &  4 days & 37 years & June 11, 2010 \\
\hline
  100,000,000,000,001 & 1024 &  5 days & 40 years & June 7, 2010 \\
\hline
  999,999,999,999,993 &  288 & 13 days & 248 years & July 2, 2010 \\
\hline
1,000,000,000,000,001 &  256 & 25 days & 283 years & July 6, 2010 \\
\hline
1,999,999,999,999,993 &  288 & 23 days & 582 years & July 29, 2010 \\
\hline
1,999,999,999,999,997 &  288 & 23 days & 503 years & July 25, 2010 \\
\hline
\end{tabular}
\caption{The running time used in each computation.}
\label{tab-time}
\end{center}
\end{minipage}
\end{table}

\section{The BBP Digit-Extraction Algorithm}\label{sect-bbp algorithm}
We briefly describe the BBP digit-extraction algorithm in this section
(see \cite{Bailey1997} for more details).
Any BBP-type formula,
such as equation (\ref{eqn-bbp}) or equation (\ref{eqn-bellard}),
can be used in the algorithm.
For simplicity,
we discuss the algorithm with equation (\ref{eqn-bbp}) in this section.

In order to obtain the $(n+1)^{\text{th}}$ bit,
compute $\FRACTION{2^n\pi}$,
where 
\[
\FRACTION{x}\DEF{=}x-\lfloor x\rfloor
\]
denotes the fraction part of $x$.
By equation (\ref{eqn-bbp}),
we have
\begin{equation}\label{eqn-2^n pi}
\FRACTION{2^n\pi}
= \FRACTION{\FRACTION{\sum_{k=0}^\infty\frac{2^{n+2-4k}}{8k+1}}
            - \FRACTION{\sum_{k=0}^\infty\frac{2^{n-1-4k}}{2k+1}}
            - \FRACTION{\sum_{k=0}^\infty\frac{2^{n  -4k}}{8k+5}}
            - \FRACTION{\sum_{k=0}^\infty\frac{2^{n-1-4k}}{4k+3}}}.
\end{equation}
Then,
evaluate each sum as below,
\begin{equation}\label{eqn-sum}
\FRACTION{\sum_{k=0}^\infty\frac{2^{n+x-4k}}{yk+z}}
= \FRACTION{
  \sum_{0\leq k<\frac{n+x}{4}}A_k
+ \sum_{\frac{n+x}{4}\leq k}B_k
},
\end{equation}
where
\begin{equation}\label{eqn-A_k,B_k}
A_k\DEF{=}\frac{2^{n+x-4k}\bmod (y k+z)}{y k+z}
\qquad\text{and}\qquad
B_k\DEF{=}\frac{1}{2^{4k-n-x}(y k+z)}.
\end{equation}
The number of terms in the first sum of equation (\ref{eqn-sum})
is linear to $n$.
Each term is a modular exponentiation followed by a floating point division.
In the second sum,
it is only required to evaluate the first $O(p/\log p)$ terms
such that $B_k>\varepsilon=2^{-p-1}$
(see equation (\ref{eqn-ulp}))
when working on $p$-bit precision.
Each term is a reciprocal computation.
For all the terms in both sums,
all the operands are integers with $O(\log n)$ bits.

The running time of the algorithm is 
\begin{equation}\label{eqn-bbp running time}
O(p(n^{1+\epsilon}+p))
\end{equation}
bit operations for any $\epsilon>0$.
Note that $p$ is required to be $\Omega(\log n)$ due to rounding error;
see Section \ref{sect-Rounding Error}.
When the algorithm is used to compute the $n^{\text{th}}$ bit
with a small $p$,
the running time is essentially linear in $n$.
However,
when the algorithm is used to compute the first $p$ bits
(i.e.\ $n=0$),
the running time is quadratic in $p$.
In this case,
there are faster algorithms
\cite{Brent1976, Salamin1976} and \cite{Chudnovsky1989},
which run in essentially linear time.

It is easy to see that the required space for the BBP algorithm is
\begin{equation}\label{eqn-bbp space}
O(p+\log n)
\end{equation}
bits.
For $p$ small,
the computation task is CPU-intensive but not data-intensive.

The algorithm is embarrassingly parallel
because it mainly evaluates summations with a large number of terms.
Evaluating these summations can be computed in parallel
with little additional overhead.

\subsection{Rounding Error}\label{sect-Rounding Error}
Since the outputs of the BBP algorithm are the exact bits of $\pi$,
it is important to understand the rounding errors that arose in the computation
and how they impact the results.
One simple way for diminishing the rounding error effect
is to increase the precision in the computation.
In practice,
at least two independent computations at different bit positions,
usually $n$ and $n-4$,
are performed in order to verify the results.
For example,
the bit sequence shown in Table \ref{tab-2e15 pi bits} was obtained
by two computations shown at the last two rows of Table \ref{tab-time}.
We discuss rounding error in more details in the rest of the section.

When a real number is represented in $p$-bit precision,
the absolute relative rounding error is bounded above by
\begin{equation}\label{eqn-ulp}
\varepsilon=0.5 \text{ {\tt ulp}}=2^{-p-1},
\end{equation}
where {\tt ulp} is the unit in the last place \cite{Goldberg1991}.
For computing the $n^{\text{th}}$ bit of $\pi$ with precision $p$,
the number of terms in the summations is $m=O(n+p)$.
The cumulative absolute relative error is bounded above by $m\varepsilon$.
For example,
when computing the $(10^{15})^{\text{th}}$ bit of $\pi$
with IEEE 754 64-bit floating point,
we have $m\approx 7\cdot 10^{14}$ (see equation (\ref{eqn-bellard})),
$p=52$ and $\varepsilon=2^{-53}$.
Then, 
$m\varepsilon\approx 0.0777>2^{-4}$,
which means that even the third bit may be incorrect due to rounding error.
In practice,
around 28 bits are calculated correctly in this case.

The long correct bit sequence can be explained
by analyzing rounding errors with a probability model as follows.
Let $\varepsilon_k$ be the error in the $k^{\text{th}}$ term
and $E=\sum \varepsilon_k$ be the error of the sum.
Suppose each $\varepsilon_k$ follows a uniform distribution
over the closed interval $[-\varepsilon,\varepsilon]$,
\[
\varepsilon_k\sim U(-\varepsilon,\varepsilon).
\]
Then, $E$ follows a uniform sum distribution
(a.k.a.\ Irwin-Hall distribution)
with mean 0 and variance $\sigma^2=m\varepsilon^2/3$.
The random variables $\varepsilon_k$'s are independent, identically distributed
and $m$ is large.
By the Central Limit Theorem,
the sum distribution can be approximated by a normal distribution
with the same mean and variance,
i.e.
\[
E\sim N(0,m\varepsilon^2/3).
\]
For $m\approx 7\cdot 10^{14}$ and $\varepsilon=2^{-53}$,
we have 
$72.79\%$ confidence of $|E|<2^{-29}$,
$97.20\%$ confidence of $|E|<2^{-28}$
and $99.999\,989\%$ confidence of $|E|<2^{-27}$.

Note that $|E|<2^{-b-1}$ does not imply $b$ correct bits
because it is possible to have consecutive 0's or 1's affected by the error.
For example,
we have used 64-bit floating point to compute bits
starting at the $(10^{15}+53)^{\text{rd}}$ position
and obtained the following 52 bits.
\begin{verbatim}
Position:  53   57   61   65   69   73   77   81   85   89   93   97   101
Hex     :  D    3    6    1    1    6    F    A    8    5    8    1    A
Binary  :  1101 0011 0110 0001 0001 0110 1111 1010 1000 0101 1000 0001 1010
                                       ^ ^^^^
\end{verbatim}
The corresponding true bit values are shown below.
\begin{verbatim}
Position:  53   57   61   65   69   73   77   81   85   89   93   97   101
Hex     :  D    3    6    1    1    7    0    9    9    E    4    3    6
Binary  :  1101 0011 0110 0001 0001 0111 0000 1001 1001 1110 0100 0011 0110
                                       ^ ^^^^
\end{verbatim}
We have $2^{-29}<|E|<2^{-28}$ but only the first 23 bits are correct
due to the rounding error at the last of the four consecutive 0's
in the true bit values.
\section{MapReduce}\label{sect-implementation}
In this section,
we discuss our Hadoop MapReduce implementation of the BBP algorithm.
For computing the bits of $\pi$ starting at position $n$ with precision $p$,
the algorithm basically evaluates the sum
\[
S=\sum_{i\in I} T_i,
\]
where each term $T_i$ consists of a few arithmetic operations;
see Section (\ref{sect-terms}).
We consider $p$ is small,
i.e.\ $p=O(\log n)$,
throughout this section.
Then,
the size of the index set $I$ is roughly $0.7n$;
see equation (\ref{eqn-bellard}).
For $n=10^{15}$,
the size of $I$ is approximately $7\cdot10^{14}$.

A straightforward approach is to partition the index set $I$
into $m$ pairwise disjoint sets $I_1,\cdots,I_m$.
Then,
evaluate each summation $\sigma_j\DEF{=}\sum_{i\in I_j} T_i$
by a mapper
and compute the final sum $S=\sum_{1\leq j\leq m}\sigma_j$
by a reducer.
However,
such implementation,
which mainly relies on map-side computation,
does not utilize a cluster
because a cluster usually has a fixed ratio
between map and reduce task capacities.
Most of the reduce slots are not used in this case.
The second problem is
that the MapReduce job possibly runs for a long time;
see Table \ref{tab-time}.
It is desirable to have a mechanism to persist the intermediate results,
so that the computation is interruptible and resumable.

In our design,
we have multi-level partitioning.
As before,
the summation is first partitioned into $m$ smaller summations
$\Sigma_j=\sum_{i\in I_j}T_i$
such that the value of $m$ is also small.
Each $\Sigma_j$ is computed by an individual MapReduce job.
A controller program executed on a gateway machine is responsible
for submitting these jobs.
The summations $\Sigma_j$ are further partitioned into tiny summations
$\sigma_{j,k}=\sum_{i\in I_{j,k}}T_i$,
where $\{I_{j,1},\cdots,I_{j,m_j}\}$ is a partition of $I_j$.
Each $\Sigma_j$ job has $m_j$ tiny summations,
which can be computed on either the map-side or the reduce-side;
see Section \ref{sect-map,reduce-side} below.
Each tiny summation task is then assigned to a node machine
by the MapReduce system.
In the task level,
if there are more than one available CPU cores in the node machine,
the tiny summation is partitioned again
so that each part is executed by a separated thread.
The task outputs $\sigma_{j,k}$'s are written to HDFS,
a persistent storage of the Hadoop cluster \cite{shv2010}.
Then,
the controller program reads $\sigma_{j,k}$'s from HDFS,
compute $\Sigma_j=\sum_{1\leq k\leq m_j}\sigma_{j,k}$
and write $\Sigma_j$ back to HDFS.
These intermediate results are persisted in HDFS.
Therefore,
the computation can possibly be interrupted at any time
by killing the controller program and all the running jobs,
and then be resumed in the future.
The final sum $S=\sum_{1\leq j\leq m}\Sigma_j$
can be efficiently computed because $m$ is relatively small.
The multi-level partitioning is summarized below.
\begin{align*}
& \text{Final Sum:} \hspace*{-3cm}
& S &= \sum_{1\leq j\leq m}\Sigma_j
\\
& \text{Jobs:}  \hspace*{-3cm}
& \Sigma_j &= \sum_{1\leq k\leq m_j}\sigma_{j,k}
\\
& \text{Tasks:}  \hspace*{-3cm}
& \sigma_{j,k} &= \sum_{1\leq t\leq m_{j,k}}s_{j,k,t}
\\
& \text{Threads:}  \hspace*{-3cm}
& s_{j,k,t} &= \sum_{i\in I_{j,k,t}}T_i
\end{align*}

\subsection{Map-side \& Reduce-side Computations}\label{sect-map,reduce-side}
In order to utilize the cluster resources,
we have developed a general framework
to execute computation tasks on either the map-side or the reduce-side.
Applications only have to define two functions:
\begin{enumerate}
\item
$\PARTITION(c, m)$:
partition the computation $c$ into $m$ parts $c_1,\dots,c_m$;
\item
$\COMPUTE(c)$:
execute the computation $c$.
\end{enumerate}

A map-side job contains multiple mappers and zero reducers.
The input computation $c$ is partitioned into $m$ parts
by a \PartitionInputFormat\ 
and then each part is executed by a mapper.
See Figure \ref{fig-map-side} below.
\begin{figure}[h]
\includegraphics[width=\textwidth, bb=0 0 1552 328]{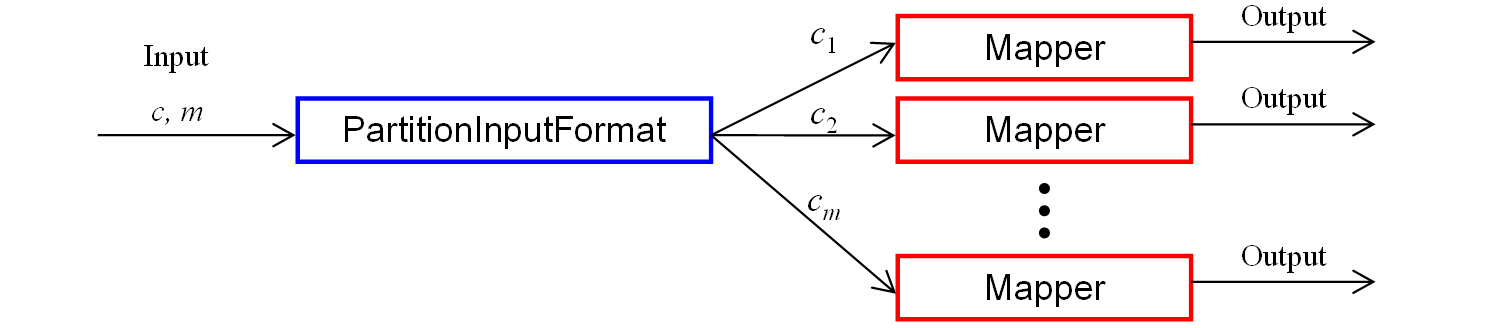}
\caption{Map-side computation.}
\label{fig-map-side}
\end{figure}

In contrast,
a reduce-side job contains a single mapper and multiple reducers.
A \SingletonInputFormat\ is used to launch a single \PartitionMapper,
which is responsible to partition the computation $c$ into $m$ parts.
Then,
the parts are forwarded to an \Indexer,
which creates indexes for launching $m$ reducers.
See Figure \ref{fig-reduce-side} below.
\begin{figure}[h]
\includegraphics[width=\textwidth, bb=0 0 1552 306]{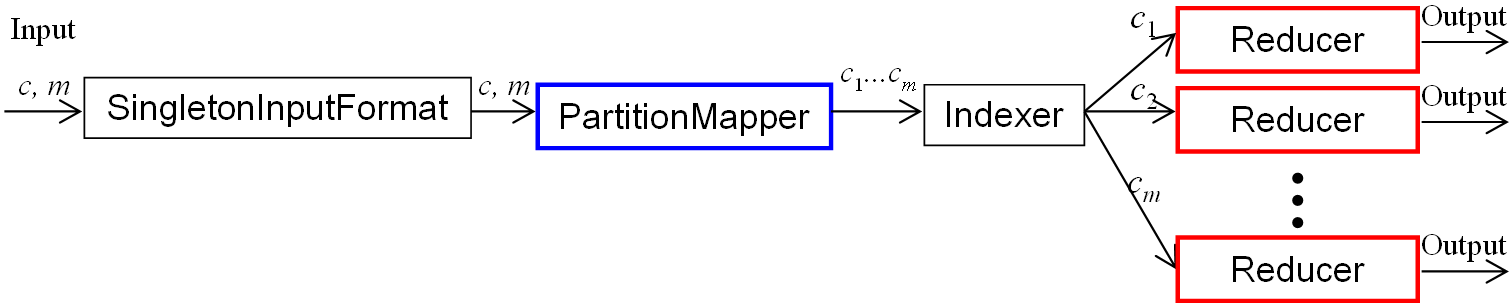}
\caption{Reduce-side computation.}
\label{fig-reduce-side}
\end{figure}

Note that the utility classes
\PartitionInputFormat,
\Mapper,
\SingletonInputFormat,
\PartitionMapper,
\Indexer\ and \Reducer\ 
are provided by our framework.
For map-side (or reduce-side) jobs,
the user defined functions $\PARTITION(c, m)$ and $\COMPUTE(c)$ are executed
in \PartitionInputFormat\ (or \PartitionMapper)
and \Mapper\ (or \Reducer),
respectively.

The map and reduce task slots in a Hadoop cluster are statically configured.
This framework allows computations utilizing both map and reduce task slots.

\subsection{Evaluating The Terms}\label{sect-terms}
As shown in equations (\ref{eqn-sum}) and (\ref{eqn-A_k,B_k}),
there are two types of terms,
$A_k$ and $B_k$,
in the summations of the BBP algorithm.
The terms $A_k$ involve a modular exponentiation
and a floating point division.
Modular exponentiation can be evaluated by the successive squaring method.
When the modulus is large,
we use Montgomery method \cite{Montgomery1985},
which is faster than successive squaring in this case.

Floating point division is implemented with arbitrary precision
because of the rounding error issue discussed
in Section \ref{sect-Rounding Error}.
For the terms $B_k$ in equation (\ref{eqn-A_k,B_k}),
the division first is done first by shifting $b=4k-n-x$ bits
and then followed by floating point division with $(p-b)$-bit precision,
where $p$ is the selected precision.

\subsection{Utilizing Cluster Idle Slices}
One of our goals is to utilize the idle slices in a cluster.
The controller program mentioned previously also monitors the cluster status.
When there are sufficient available map (or reduce) slots,
the controller program submits a map-side (or reduce-side) job.
Each job is small so that it holds cluster resource only
for a short period of time.

In one of our computations (see the last row in Table \ref{tab-time}),
we had $n=2\cdot10^{15}-3$ and $p=288$.
The summation had approximately $1.4\cdot10^{15}$ terms.
It was executed in a 1000-node cluster.
Each node had two quad-core CPUs
with clock speed ranging from 2.0 GHz to 2.5 GHz,
and was configured to support four map tasks and two reduce tasks.
The computation was divided into 35,000 jobs.
Depending on the cluster load condition,
the controller program might submit up to 60 concurrent jobs at any time.
A job had 200 mappers with one thread each
or 100 reducers with two threads each.
Each thread computed a summation with roughly 200,000,000 terms
and took around 45 minutes.
The entire computation took 23 days of real time and 503 years of CPU time.
\section{Conclusions \& Future Works}\label{sect-future}
In this paper, we present our latest results on computing $\pi$ using Apache
Hadoop. We extend the previous record of calculating specific bits of $\pi$
from position around the one quadrillionth bit to 
position around the two quadrillionth bit,
and from 64-bit precision to 256-bit precision.
The distributed
computation is done through a MapReduce program called DistBbp. 
Our elastic computation framework automatically schedules computation slices as
either map-side or reduce-side computation to fully exploit idle cluster
resources.

A natural extension of this work is to compute the bits of $\pi$ at higher
positions, say the ten quadrillionth bit position, or even the quintillionth
bit position.  Besides, it is interesting to compute all the first $n$ digits
of $\pi$ with Hadoop clusters.  Such computation task is not only CPU-intensive
but also data-intensive.
\section*{Acknowledgment}
We thank Robert Chansler and Hong Tang
for providing helpful review comments and suggestions for this paper.
Owen O'Malley, Chris Douglas, Arun Murthy and Milind Bhandarkar 
have provided many useful ideas and help in developing DistBbp.
We also thank
Eric Baldeschwieler, Kazi Atif-Uz Zaman and Pei Lin Ong
for supporting this project.
\noindent
\bibliographystyle{plain}
\bibstyle{plain}
\bibliography{2e15bit_of_pi}

\end{document}